\pdfoutput=1
\documentclass[journal]{IEEEtran}

\ifCLASSINFOpdf

\else

\fi

\usepackage[utf8]{inputenc}
\usepackage[english]{babel}
\usepackage[nonumberlist,nopostdot,nogroupskip]{glossaries}
\usepackage[normalem]{ulem}
\usepackage[para,flushleft]{threeparttable}
\usepackage{wasysym} 
\usepackage{tcolorbox} 
\usepackage{tocbasic} 
\usepackage{comment}

\usepackage{cite}
\usepackage{graphicx}
\usepackage{amssymb}
\usepackage{amsmath}
\usepackage{float}
\usepackage{array}
\usepackage{multirow}
\usepackage{enumerate}
\usepackage{enumitem}
\usepackage{xspace}
\usepackage{subcaption}
\usepackage{mwe}
\usepackage[hyphens]{url} 
\usepackage{cleveref}

\makeglossaries

\newacronym{4g}{4G}{fourth-generation}
\newacronym{5g}{5G}{fifth-generation}
\newacronym{nr}{NR}{New Radio}
\newacronym{lte}{LTE}{Long-Term Evolution}
\newacronym{3gpp}{3GPP}{3rd Generation Partnership Project}
\newacronym{oai}{OAI}{OpenAirInterface}
\newacronym{awgn}{AWGN}{additive white Gaussian noise}
\newacronym{snr}{SNR}{signal-to-noise ratio}
\newacronym{ml}{ML}{maximum likelihood}
\newacronym{ue}{UE}{user equipment}
\newacronym{bs}{BS}{base-station}
\newacronym{zf}{ZF}{zero-forcing}
\newacronym{mf}{MF}{matched filter}
\newacronym{tdd}{TDD}{time-division duplexing}
\newacronym{cdm}{CDM}{code-division multiplexing}
\newacronym{gnb}{gNB}{next-generation node B}
\newacronym{srs}{SRS}{sounding reference signal}
\newacronym{dmrs}{DM-RS}{demodulation reference signal}
\newacronym{pdsch}{PDSCH}{physical downlink shared channel}
\newacronym{pusch}{PUSCH}{physical uplink shared channel}
\newacronym{gui}{GUI}{graphical user interface}
\newacronym{ul}{UL}{uplink}
\newacronym{dl}{DL}{downlink}
\newacronym{cots}{COTS}{commercial off-the-shelf}
\newacronym{siso}{SISO}{single-input single-output}
\newacronym{mimo}{MIMO}{multiple-input multiple-output}
\newacronym{su}{SU}{single-user}
\newacronym{mu}{MU}{multi-user}
\newacronym{cdl}{CDL}{clustered delay line}
\newacronym{rrh}{RRH}{remote radio head}
\newacronym{bbu}{BBU}{baseband unit}
\newacronym{ran}{RAN}{radio access network}
\newacronym{phy}{PHY}{physical layer}
\newacronym{sdr}{SDR}{software-defined radio}
\newacronym{sme}{SME}{small and medium-sized enterprises}
\newacronym{harq}{HARQ}{hybrid automatic repeat request}
\newacronym{ldpc}{LDPC}{low-density parity-check}
\newacronym{fec}{FEC}{forward error correction}
\newacronym{soc}{SoC}{system~-~on~-~chip}
\newacronym{ru}{RU}{radio unit}
\newacronym{tm}{TM}{transmission mode}
\newacronym{ula}{ULA}{uniform linear array}
\newacronym{osa}{OSA}{OpenAirInterface Software Alliance}
\newacronym{cn}{CN}{core network}
\newacronym{los}{LOS}{line-of-sight}
\newacronym{nlos}{NLOS}{non-line-of-sight}
\newacronym{umi}{UMi}{urban micro}
\newacronym{ofdm}{OFDM}{orthogonal frequency-division multiplexing}
\newacronym{sse}{SSE}{Streaming SIMD Extensions}
\newacronym{avx}{AVX}{Advanced Vector Extensions}
\newacronym{simd}{SIMD}{single-instruction multiple-data}
\newacronym{gpp}{GPP}{general purpose processor}
\newacronym{scs}{SCS}{subcarrier spacing}
\newacronym{cpri}{CPRI}{Common Public Radio Interface}
\newacronym{llr}{LLR}{log-likelihood ratio}
\newacronym{dft}{DFT}{discrete Fourier transform}
\newacronym{ota}{OTA}{over-the-air}
\newacronym{rt}{RT}{real-time}
\newacronym{noma}{NOMA}{non-orthogonal multiple access}
\newacronym{sword}{SWORD}{SoftWare Open Radio Design}
\newacronym{must}{MUST}{multi-user superposition transmission}
\newacronym{ulsch}{UL-SCH}{uplink shared channel}
\newacronym{dlsch}{DL-SCH}{downlink shared channel}
\newacronym{tb}{TB}{transport block}
\newacronym{mrt}{MRT}{maximum ratio transmission}
\newacronym{pas}{PAS}{power angle spectrum}
\newacronym[\glslongpluralkey={angles-of-arrival}]{aoa}{AoA}{angle-of-arrival}
\newacronym[\glslongpluralkey={angles-of-departure}]{aod}{AoD}{angle-of-departure}
\newacronym{fft}{FFT}{fast Fourier transform}
\newacronym{se}{SE}{spectral efficiency}
\newacronym{mph}{mph}{miles per hour}
\newacronym{csi}{CSI}{channel state information}
\newacronym{usrp}{USRP}{Universal Software Radio Peripheral}
\newacronym{bios}{BIOS}{basic input/output system}
\newacronym{cpu}{CPU}{central processing unit}
\newacronym{mkl}{MKL}{Math Kernel Libary}
\newacronym{isa}{ISA}{instruction set architecture}
\newacronym{fftw}{FFTW}{Fastest Fourier Transform in the West}
\newacronym{asic}{ASIC}{Application-Specific Integrated Circuit}
\newacronym{fpga}{FPGA}{field-programmable gate array}
\newacronym{dsp}{DSP}{digital signal processing}
\newacronym{re}{RE}{resource element}
\newacronym{mcs}{MCS}{modulation and coding scheme}
\newacronym{tpc}{TPC}{transmit power control}
\newacronym{epre}{EPRE}{energy per resource element}
\newacronym{rrc}{RRC}{radio resource control}
\newacronym{dci}{DCI}{downlink control information}
\newacronym{amc}{AMC}{adaptive modulation and coding}
\newacronym{pa}{PA}{power amplifier}
\newacronym{lo}{LO}{local oscillator}
\newacronym{mmse}{MMSE}{minimum-mean-square-error}
\newacronym{nl}{NL}{non-linear}

\begin{document}

\title{Towards Radio Designs with Non-Linear Processing for Next Generation Mobile Systems}

\author{Konstantinos Nikitopoulos, Marcin Filo, Chathura Jayawardena and Rahim Tafazolli \thanks{The authors are with the
	Wireless Systems Lab, 5G Innovation Centre, Institute for Communication Systems, University of Surrey, Guildford GU2 7XH, UK e-mail: \{k.nikitopoulos, m.filo, c.jayawardena, r. tafazolli\}@surrey.ac.uk.}}
\maketitle
\begin{abstract}

MIMO mobile systems, with a large number of antennas at the base-station side, enable the concurrent transmission of multiple, spatially separated information streams and, therefore, enable improved network throughput and connectivity both in uplink and downlink transmissions. Traditionally, to efficiently facilitate such MIMO transmissions, \textit{linear} base-station processing is adopted, that translates the MIMO channel into several single-antenna channels. Still, while such approaches are relatively easy to implement, they can leave on the table a significant amount of unexploited MIMO capacity. Recently proposed \textit{non-linear} base-station processing methods claim this unexplored capacity and promise a substantially increased network throughput. Still, to the best of the authors' knowledge, non-linear base-station processing methods not only have not yet been adopted by actual systems, but have not even been evaluated in a standard-compliant framework, involving of all the necessary algorithmic modules required by a practical system. This work, outlines our experience by trying to incorporate and evaluate the gains of non-linear base-station processing in a 3GPP standard environment. We discuss the several corresponding challenges and our adopted solutions, together with their corresponding limitations. We report gains that we have managed to verify, and we also discuss remaining challenges, missing algorithmic components and future research directions that would be required towards highly efficient, future mobile systems that can efficiently exploit the gains of non-linear, base-station processing.
 
\end{abstract}

\section{Introduction}
Much of the current communication systems research focuses on finding new, breakthrough ways to increase the achievable throughout (both at a user and a system level) and user connectivity capabilities, while meeting very tight latency requirements. In this direction, a plethora of ideas have been proposed. Still, very few of these ideas, and perhaps the simplest in terms of practical realization, have finally been adopted by actual wireless systems and standards. In the natural question
``why is this happening?'' someone can give several answers. In many of the works published, the proposed ideas are only evaluated via simulations and, therefore, the results may be heavily \textit{assumption-depended}. Namely, the showed gains can be a strong function of the simulated environment that can sufficiently differ from the actual transmission environment. To facilitate more realistic evaluations, many researchers use ``proof-of-concept'' systems. However, this approach comes with its own challenges and practical limitations. Such a challenge is the availability of appropriate research platforms able to realize and validate proposed novel ideas. In addition, in many cases and especially in physical layer research, the proposed ideas are not evaluated in a complete, standard-compliant environment. As a result, \textit{additional algorithmic components} may be required to make a new idea adoptable by a practical system or communication standards. As we will discuss later in detail, such components are often related to limitations imposed from the existing system design, as for example limitations related to conventional signalling procedures or even other mechanisms that are required for transmsission optimization, as for example the transmission rate selection. As a result, implementing and evaluating new physical layer concepts and ideas in a standard-compliant environment can be of substantial importance, not only to verify the potential gains in more realistic transmission scenarios, but also to identify any need for new ``building blocks'' or required modifications in the standards (e.g., the signalling, pilots, access methods). In other words, testing and verifying new physical layer ideas in a research-grade, standard-compliant environment can be an important step towards future systems design and evaluation, that can highly increase our confidence on newly proposed approaches and can assist in identifying further requirements and missing components that will enable the adoption of novel ideas in actual systems.

In this work we focus on recently proposed ideas to improve \gls{bs}
processing in \gls{mimo} 
spatially multiplexed systems. The use of a large number of antennas at the \gls{bs} 
side has been shown to be a very efficient way to increase the achievable throughput and the user connectivity capabilities of mobile systems, both in uplink and downlink transmissions, by enabling several concurrently transmitting, spatially separated users (i.e., Multi-User \gls{mimo})
\cite{bigstation,shepard2012argos,tan2009sam}. Traditionally, in such systems, \textit{linear} precoding (in the downlink) and detection (in the uplink) approaches are employed at the \gls{bs}, 
based on the \gls{zf} or on the \gls{mmse} 
principles. Such linear approaches have two major practical benefits. Their implementation is relatively simple, and since they practically translate the mutually interfering information streams into traditional, non-interfering ones, they can be easily adopted by standards with minimum modifications. Still, their main drawback is that, in order for these approaches to be efficient in terms of achievable throughput, the number of 
\gls{bs} antennas needs to be much larger than the number of concurrently transmitted information streams, and, therefore, the number of served users \cite{shepard2012argos,bigstation}. However, since by increasing the number of antennas the capacity of the \gls{mimo} 
channel generally increases \cite{telatar1999capacity}, such an approach leaves on the table a significant amount of unexploited capacity \cite{nikitopoulos2014geosphere}. Equivalently, it unnecessarily increases the number of 
\gls{bs} antennas for a certain number of users, significantly increasing the 
\gls{bs} cost and reducing power efficiency. In contrary, \textit{non-linear} 
\gls{bs} processing approaches, like ``hard'' and ``soft'' Maximum-Likelihood detection, in the uplink \cite{nikitopoulos2014geosphere,STS}, and Vector-Perturbation in the downlink \cite{hochwald2005vector}, promise substantially increased achievable throughput and user connectivity.
Still, to the best of our knowledge, such approaches have not yet been adopted by practical systems and have not even been validated in a standard-compliant scenario. In addition, and perhaps as a consequence, it is not obvious what further system changes are required to deliver the promised gains in practice. 

\noindent\textbf{Contributions of this paper.} In this work we present our experience by trying to incorporate and validate the performance gains of advanced, non-linear methods in a \gls{3gpp} compliant framework. 
We outline the main challenges we have faced in order to realize and evaluate such approaches, together with our adopted approaches and their limitations. We report gains that we have been able to verify, and we describe missing components, remaining challenges, potential solutions, and open research directions that would enable the adoption of such approaches in practice. Finally, we discuss some future research directions that have the potential to substantially increase both the throughput and connectivity capabilities of next generation wireless systems 
by adopting new forms of non-linear 
\gls{bs} processing.





\section{Outline of Non-linear Processing Techniques for Base-Station Processing}
As discussed, current \gls{mimo} 
deployments mostly employ linear \gls{bs} 
processing, both for uplink and downlink transmissions, but such approaches may leave a significant amount of capacity unexploited.
Instead, non-linear \gls{bs} 
processing approaches promise consistent gains compared to the linear ones, both in terms of achievable throughput and user connectivity. For uplink transmissions, ``hard'' Maximum-Likelihood detection methods have been proposed, both exact and approximate, and with most of them being realized in terms of sphere decoding \cite{BurgHardSD,nikitopoulos2014geosphere,FSD,GSD,shabanykbest,Nilssonkbest}. 
Most of these approaches, though, have been evaluated using simulations and by assuming ``Rayleigh'' or other mathematically modeled \gls{mimo} 
channels. While such mathematical models are necessary for the theoretical analysis of such systems, they do not necessary capture the spatial multiplexing capabilities of the actual \gls{mimo} 
channels. In addition, the provided performance of these methods is often presented in terms of (uncoded) bit-error-rate, which is not adequate for evaluating systems throughput gains. The sphere decoding approaches of \cite{nikitopoulos2014geosphere,FlexCore,FSD} are evaluated in actual transmission channel environments and in terms of achievable throughput. Still, their evaluations are based on a very limited number of transmission rates (i.e., combinations of QAM constellation size and coding rates) that are selected based on their average performance. In addition, since the processing takes place off-line, the reported achievable rates do not include the impact of the higher layers of the protocol stack (e.g., the impact of the \gls{harq} mechanism). In addition, ``hard'' detection cannot be used jointly with state-of-the-art ``soft'' channel encoding and decoding schemes (e.g., LDPC, Turbo) adopted in recent standards, and therefore, are of limited practical interest.
 For use with soft channel decoding schemes, soft-output sphere decoders have been proposed to reduce the complexity of optimal soft detection \cite{STS,Tuplesoftout,SFSD,jalden2005parallel}. 
 Again, most soft sphere decoding approaches, including the sequential sphere decoder of \cite{STS} and the soft fixed complexity sphere decoder of \cite{SFSD}, are evaluated by assuming mathematically modelled \gls{mimo} channels. The massively parallel hard and soft detectors of \cite{nikitopoulos2018massively,5GRefWileyKN}, that enable practical low complexity and low latency non-linear detection, are evaluated both in mathematically modelled and measured channels. Still, these evaluations are based on a limited number of transmission rates, and the performance is reported for rates that have been chosen by an exhaustive search to maximize the average throughput, across all positions, instead of optimizing rate per packet transmission. In addition, similarly to the hard detection approaches, they do not capture the impact of the higher layers of the protocol stack.
 

In the downlink, non-linear, theoretical precoding approaches exist which claim the \gls{mimo} 
channel capacity that is currently unexploited by linear precoders \cite{hochwald2005vector,FCSE,MMSEVP,gain15dbvp,improvedvp,VPgeneralized}. 
These approaches are based on Dirty Paper Coding principles which can achieve the capacity of the Gaussian broadcast channel \cite{costa1983writing}. In this direction, the non-linear Tomlinson-Harashima precoding \cite{harashima1972matched} can substantially improve on the throughput achievable by traditional linear precoding. 
Improving on the Tomlinson-Harashima precoding, vector perturbation \cite{hochwald2005vector} precoding can further contribute into bridging the gap to the \gls{mimo} 
channel capacity limit. In particular, efficiently perturbing the transmitted constellation symbols in a way that the corresponding perturbation effect can be efficiently compensated at the receiver side. 
Again, most evaluations of vector perturbation precoding \cite{hochwald2005vector,FCSE} are limited to simulations employing mathematically modelled channels. The massively parallel vector perturbation precoder of \cite{husmann2018viper}, that promises practical non-linear precoding, is evaluated by over-the-air experiments, but also with off-line processing, inheriting the corresponding evaluation drawbacks.

As discussed, to the best of our knowledge, none of the above approaches has been evaluated in a standard-compliant framework, neither a corresponding attempt has been reported that would identify missing algorithmic components, and further challenges that need to be resolved.


\section{Challenges, Adopted Approaches \& Lessons Learned}

Here we describe our experience by trying to incorporate and validate the performance of non-linear processing approaches in a 3GPP compliant environment. We outline some of the main challenges we have faced, as well as our adopted solutions together with their corresponding limitations, and the related lessons we have learned. As we will discuss in detail, such an attempt came with numerous challenges, ranging from finding (and extending) an appropriate software and hardware platform to perform our evaluations, to challenges related to missing components and practical aspects of the algorithms that, to the best of our knowledge, have not been highlighted/identified before.




\subsection{Seeking for the appropriate software platform}

There is a number of software platforms which aim at providing a \gls{3gpp} compliant protocol stack, capable of running on general-purpose processors. They can be broadly classified as commercial and open-source. The commercial solutions include, among others, the LTE and NR Network Software Suit by Amarisoft \cite{Amarisoft}, the National Instruments LTE Application Framework for LabVIEW Communications System Design Suite \cite{LTEAPPFW} and Intel's FlexRAN \cite{FlexRAN}. The most complete is perhaps the solution provided by Amarisoft which, in contrast to other options, provides a full protocol stack implementation on \gls{bs} side and \gls{ue} side. Although it supports many features and transmission modes, due to its closed-source nature, Amarisoft solution cannot be openly used for physical layer research. In contrast to commercial platforms, open-source solutions which include srsLTE \cite{srsLTE}, openLTE \cite{openLTE} and \gls{oai} \cite{kaltenberger2020openairinterface} are freely available to the public. Among those, it seems that the most advanced platform is \gls{oai} which is the open-source solution with the largest developer community actively working towards adding new features into the existing code base (e.g. support for 5G \gls{nr}).

\noindent\textbf{Our adopted solution.} For our evaluations, we extended our recently proposed \gls{sword} platform \cite{8938716}, that overcomes the missing support for large/massive \gls{mimo} setups, as well as the inherent inability of existing approaches to investigate non-linear processing without prohibitive software and hardware optimization necessary. 
To support downlink and uplink MU-MIMO transmission schemes, which were in our main interest for testing non-linear processing approaches, \gls{sword} significantly extends the \gls{oai} code base \cite{9053352} and introduces a completely new mode of operation which we call \textit{pseudo-\gls{rt}}. This new mode combines the properties and builds upon two existing modes of operations already supported by the \gls{oai} which permit \gls{rt} \gls{ota} transmission and emulation of an entire radio access network without the use of \gls{sdr} modules. Compared to the generally adopted method of the \textit{offline processing} in which a received signal is stored in a raw format on the receiver side and then processed, the \textit{pseudo-\gls{rt}} can be effectively used to evaluate the impact of advanced physical layer approaches on the overall system performance. In contrast to the \textit{offline processing}, the pseudo-\gls{rt} mode makes use of a pause period between each transmission to facilitate signal processing on both sides. As a result, it preserves the dependence between consecutive events, allowing for a more realistic setting in which the full-protocol stack is executed.

In order to enable pseudo-\gls{rt} processing, a mechanism is required to ensure that processing of a subframe at a receiver side begins only after processing at a transmitter side is completed. For this purpose, the \textit{subframe processing synchronization mechanism} developed as part of the \gls{oai} emulation mode was reused and extended. The extensions include a number of changes in the \gls{oai} device library responsible for handling communication with \gls{sdr} modules to allow for appropriate handling of multiple \glspl{ue} and multiple \gls{bs} radio chains. Further, to account for various aspects related to transmission over a real channel (e.g. propagation delay), a subset of routines used in \gls{oai} \gls{rt} mode were also modified. It is worth noting that the \textit{subframe processing synchronization mechanism} does not ensure the exact time when transmission and reception is initiated on each \gls{sdr} module. To circumvent this, additional synchronization between \gls{sdr} modules is required. This can be achieved with an external reference clock source which is common for all \gls{sdr} modules on the \gls{bs} side and \gls{ue} side.

\noindent\textbf{Lessons learned.} 
The effective investigation of advanced physical layer approaches requires supporting large/massive \gls{mimo} setups and pseudo-\gls{rt} mode of operation, which are not yet available in existing platforms. 
While our platform provides these features, the current implementation of the pseudo-\gls{rt} mode of operation mandates that processing for all \glspl{ue} and \gls{bs} is preformed by a single process, 
executed on a single workstation. Although beneficial during development and debugging of new features, we found that this architecture does not scale well for a higher number of \glspl{ue} and \glspl{bs} due to limited computational power. In the next iteration of our software platform we intend to adopt a new architecture which permits \gls{ue} processing to be executed in a separate process (and a separate machine) to allow for better flexibility in allocation of resources for processing. Note that this is also a key enabler in providing more flexibility in interconnecting \gls{sdr} modules, as the current software architecture mandates that all radio modules are connected to the same workstation. As a result, in order to conduct measurements under various channel conditions, long, low-attenuation cables are required which interconnect \gls{ue} antennas with \gls{sdr} modules on the \gls{ue} side. We noticed that these cables, due to their limited length, can significantly restrict the set of scenarios that can be investigated. To address this we foresee to rework the \textit{subframe processing synchronization mechanism} which constitutes the core of the pseudo-\gls{rt} mode, and thus eliminate the need for such cables. Given the new architecture, the reworked mechanism would allow for the flexibility in interconnecting \gls{sdr} modules used on the \gls{ue} side with any workstation dedicated for \gls{ue} processing.

\subsection{Seeking for the appropriate hardware platform.}

There is a number of hardware platforms capable of supporting \gls{mimo} setups which aim to be open to everyone for experimentation and can be potentially used for evaluation of new physical layer solutions. One example of such a hardware platform is COSMOS \cite{COSMOS} which is a city-scale testbed deployed in the city of New York aimed at providing means for real-world experimentation on next generation wireless technologies and applications. Another example is POWDER \cite{POWDER} which is also a cite-scale testbed run by the University of Utah. Contrary to COSMOS, POWDER provides hardware components specifically dedicated for large/massive MIMO experimentation, with up to 64 antennas per site/sector. Interestingly, both COSMOS and POWDER allow for the use of various open-source software platforms such as \gls{oai}, srsLTE, or openLTE. Yet another example of a hardware platform is LuMaMi \cite{malkowsky2017world} of Lund University. LuMaMi is much smaller in scale compared to COSMOS and POWDER, but in contrast to the other two testbeds, it is specifically dedicated for conducting large/massive MIMO related research and supports up to 128 radio chains. 

Although all three setups have a broad range of capabilities, they come with certain limitations, that make them non appropriate for meeting our objectives, at least at their current design stage. For instance, in case of COSMOS, the capabilities of \gls{sdr} modules used in the deployed nodes are limited to a maximum of four radio chains per site/sector. This can be potentially circumvented by considering a distributed MIMO setup, however, due to additional challenges, this type of setup currently is not our main focus.
Situation is slightly different in case of POWDER. In this case the limitation resides on the \gls{ue} side, as only two \gls{sdr} modules in POWDER's massive MIMO setup seems to be currently dedicated to run as \glspl{ue}. This means that the non-linear processing gains would be difficult to demonstrate since they target supporting numbers of users that is similar to the numbers of \gls{bs} antennas \cite {nikitopoulos2014geosphere,FlexCore,husmann2018viper}.
The main limitation of LuMaMi is that, contrary to the other testbeds, it heavily relies on proprietary hardware and software solutions from National Instruments \cite{NIMIMO}. This means that any experiments would have to be based on National Instruments' software. Note that LuMaMi was not designed to be used for evaluation of physical layer approaches as part of a full \gls{3gpp} compliant protocol stack and it is not clear if LuMaMi would support the National Instruments software extensions which could potentially bridge this gap.
In addition to the above, in all three cases, the lack of physical access to nodes dedicated for experimentation on the \gls{ue} side, restricts investigation to a limited set of scenarios.

\noindent\textbf{Our adopted solution.}
The identified limitations of the existing publicly available hardware platforms convinced us to invest in development of our own hardware platform which can be easily moved around and which permits investigation of scenarios with different number of \gls{bs} antennas, and different number of \glspl{ue}.
The main hardware component of our \gls{sword} hardware platform is a multi-core x86\_64 workstation with a large number of PCIe lanes. The large number of PCIe lanes is necessary to host multiple NICs, which in turn we use to interface with \gls{sdr} modules of our choice. The \gls{sdr} module selected is the \gls{usrp} X series with UBX daughterboard. \gls{usrp} X series hosts two independent radio chains and is one of the \gls{sdr} modules recommended by Ettus for applications which require phase alignment \cite{USRP_UBX}. In order to synchronize and maintain phase alignment across multiple \gls{sdr} modules we exploit the Ettus Research Octoclock-G CDA-2990 \cite{octoclock} which is a highly-accurate external clock reference and pulse distribution module. The synchronized USRPs on the \gls{bs} side are connected to a 3.4-3.8GHz antenna array which is composed of 64 half wavelength-spaced elements in a dual-polarized, 8x8 configuration. Circulators are used to connect TX and RX paths of each radio chain to an antenna port. More detailed description with a rational behind using specific building blocks can be found in \cite{8938716}.

\noindent\textbf{Lessons learned.}
 In order to investigate and demonstrate the benefits of non-linear processing, a movable hardware platform which can run multiple \glspl{ue} and support large~/~massive \gls{mimo} setup is needed. While our \gls{sword} hardware solution meets these requirements, maintaining phase alignment through reference clock sharing across multiple \glspl{usrp} proved to be difficult and required frequent execution of \gls{tdd} reciprocity calibration to compensate for any drifts. We observed that such drifts had a significant negative impact on system performance, in particular when number of \glspl{ue} in a setup was approaching number of \gls{bs} antennas. In order to improve this we plan to achieve phase alignment through the \gls{lo} sharing, rather than reference clock sharing. As highlighted by Ettus in \cite{NI_RFSYNC}, the \gls{lo} sharing can significantly reduce short term and long term phase drift. Note that \gls{usrp} N32X series would be required for this purpose. Further, our existing hardware setup is currently based on the use of circulators which, due to the limited output power of \gls{usrp} X series, significantly limits the range of scenarios which can be investigated. In order to overcome this we intend to replace circulators with external power amplifiers in the next iteration of our platform.

\subsection{Remaining system challenges and tweaks around them}
\label{remaining_challenges}
While trying to evaluate the non-linear approaches we came across a 
number of practical issues that needed either to be resolved or to bypassed. These are:

\noindent\textbf{Enabling non-linear processing.}
As many non-linear decoding approaches are designed for OFDM transmission, in order to test non-linear processing in the uplink, we modified the processing of \gls{pusch} in our LTE-based platform by making transform precoding optional (see LTE \gls{pusch} processing in \cite{3gpp.36.211}). To inform \glspl{ue} about the use of transform precoding we extended \gls{rrc} signaling in line with the 5G-NR specification (note that transform precoding in 5G-NR is optional and can be dynamically enabled or disabled using \gls{rrc} signaling). We faced similar issues with non-linear precoding approaches in the downlink which adopt vector perturbation and thus require a modulo operation to be applied at the transmitter side \cite{hochwald2005vector,improvedvp}.  
To revert this operation on the receiver side, we modified \gls{ue} processing of \gls{pdsch} accordingly. Furthermore, we extended \gls{rrc} signaling to inform \glspl{ue} about the use of vector perturbation. Note that to enable more dynamic switching 
existing set of \glspl{dci} in LTE and 5G-NR used for scheduling transmission opportunities could be extended to include information on whether the incoming transmission underwent vector perturbation.

\begin{figure}
  \includegraphics[width=\columnwidth]{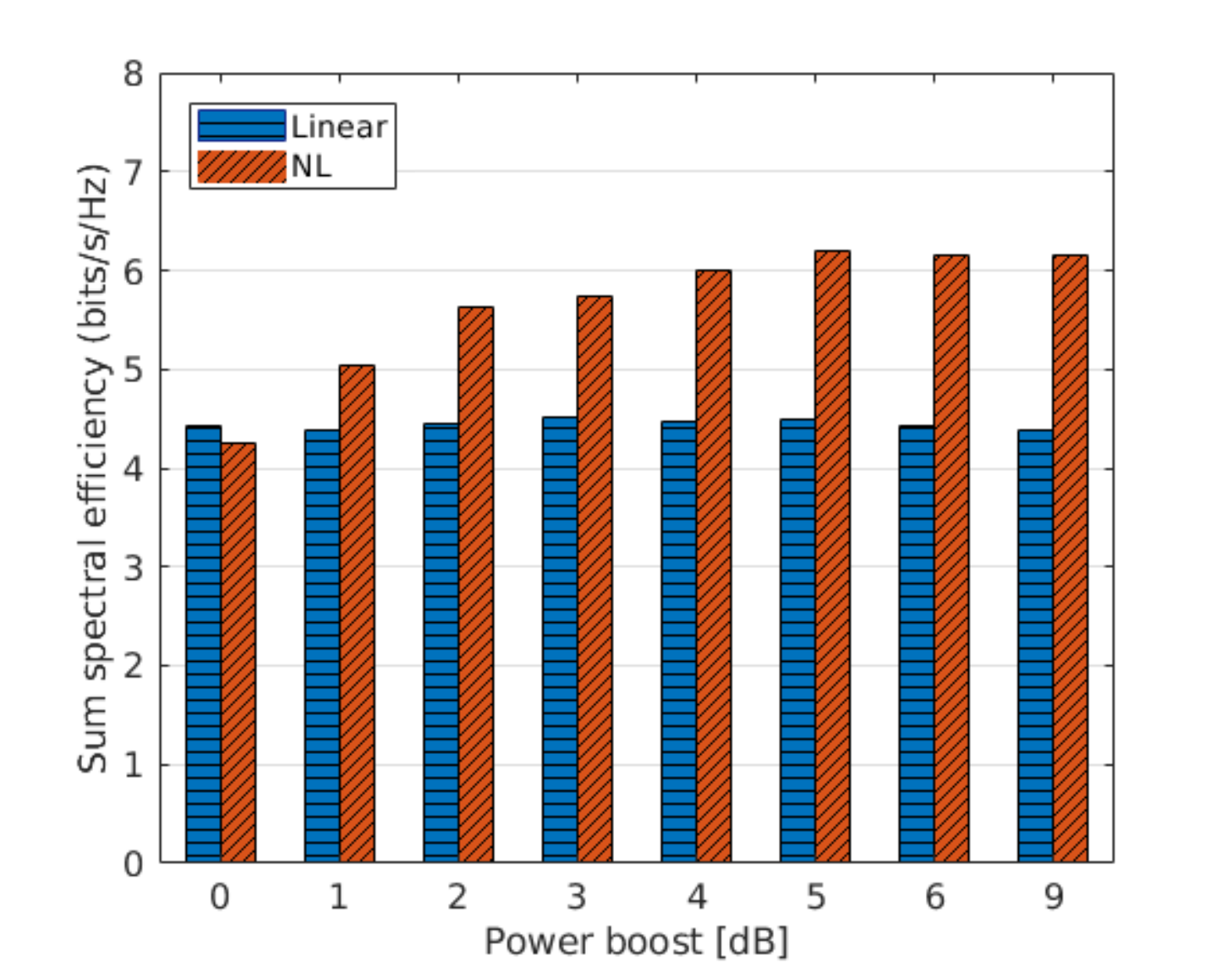}
  \caption{The impact of \gls{ue} specific \gls{dmrs} power boosting on system performance.}
  \label{fig.pilot_boost}
  \vspace{-0.4cm}
 \end{figure}

\noindent\textbf{Transmission rate selection.} \Gls{amc} is an important aspect of \gls{3gpp} systems that enables the efficient utilization of the available spectrum resources. However, \Gls{amc} for non-linear is still an open problem.  As discussed before, in order to evaluate performance of non-linear algorithms, the research community usually conducts an exhaustive search by running experiments for a small number of rates (i.e., QAM constellations and channel coding rates) and shows the average performance per rate. Although useful, the number of rates is in general very limited. In order to better evaluate the performance of non-linear approaches, and in the absence of \Gls{amc}, we have applied an ``adaptive'' rate adaptation algorithm which selects the employed \gls{mcs} based on the reported ACK/NACK information. More specifically, the employed algorithm tracks the erroneous and correctly received transmissions in both uplink and downlink. Based on this information, and given a maximum and minimum \gls{mcs}, the algorithm attempts to adjust the \gls{mcs} value after a predefined number of consecutive ACKs, or NACKs is received (resetting a ACK counter, when a NACK is received, and NACK counter, when an ACK is received). To prevent excessive \gls{mcs} switching, the proposed \gls{mcs} selection approach exploits also a simple ``cool-off period mechanism'' that prevents any \gls{mcs} changes for a specific number of frames after the last \gls{mcs} change. Still, while our adopted approach can provide an improved throughput evaluation compared to traditionally used approaches that use a limited number of rates (and they depict the rate that maximizes the average performance across channels), it is far from being realistic, and can only be used to reliably evaluate the performance in a static environment where channel remains relatively unchanged over multiple radio frames.
\noindent\textbf{DL channel estimation.}
As indicated in \cite{hochwald2005vector} non-linear precoding approaches, such as vector perturbation, results in a higher transmitted symbol constellation.
As we have here identified this makes non-linear approaches more sensitive to the channel estimation errors than linear approaches.
In order to evaluate the performance of non-linear precoding approaches adopting vector-perturbation, we compensated for the impact of the channel estimation errors by boosting the transmit power of \gls{ue} specific \gls{dmrs} used in LTE and 5G-NR for channel estimation. We note that \gls{dmrs} is only used for detection at the \gls{ue} side and not for beamforming that is based on the \gls{srs}. We also note that LTE and 5G-NR already support power boosting for \gls{ue} specific \gls{dmrs}, however, only a predefined set of power boosting values can be used for this purpose. In case of LTE, a 3dB power boosting is used when more than 2 layers are transmitted. In case of 5G-NR 3dB, or 4.77dB power boosting can be applied, depending on the \gls{dmrs} configuration used \cite{3gpp.38.214}. To inform \glspl{ue} about the non-standard compliant values we also extended \gls{rrc} signaling. Figure \ref{fig.pilot_boost} presents data for a single indoor measurement position and depicts the impact of \gls{ue} specific \gls{dmrs} power boosting on downlink sum spectral efficiency for a 4x4 MU-MIMO configuration. As seen, power boosting of \gls{ue} specific \gls{dmrs} can lead to significant performance improvements when non-linear precoding is used. 

\noindent\textbf{Channel state information (CSI) estimation.}
Another issue that we came across, independent of processing type (i.e. linear or non-linear), is related to estimation of \gls{csi}. 
For the purpose of \gls{csi} estimation 5G-NR and LTE employ a special signal transmitted in uplink termed \gls{srs}. Existing implementation of \gls{srs} \gls{tpc} mechanism may result, however, in a partial loss of \gls{csi}, which in turn can limit performance of a precoder. In particular, the signal amplitude difference between multiple \glspl{ue} in a cell is lost. 
The fundamental objective of the \gls{tpc} mechanism is to assure that signals transmitted by multiple \glspl{ue} arrive at BS with approximately the same strength, which in turn results in a mentioned loss of information. 
In order to circumvent this, as a first attempt solution, we set the \gls{srs} transmit power to a constant value. To achieve it, and at the same time retain the benefits of \gls{tpc} for uplink, we separated \gls{tpc} for \gls{srs} and other uplink signals so that they are not conducted jointly. Note that in 5G-NR separate \gls{tpc} for \gls{srs} and other uplink signals is already part of the standard. Separate \gls{tpc} for a new variant of \gls{srs} (termed ``additional'' \gls{srs}) has been also recently introduced in LTE release 16.

\noindent\textbf{Lessons learned.} Adapting non-linear processing in real system requires a number of changes in \gls{3gpp} standards which include changes in \gls{pusch} and \gls{pdsch} processing. Additional changes are also required in the signaling procedures. These primarily include extensions of \gls{rrc} signaling which is used to inform \glspl{ue} about the settings of non-linear processing (e.g. additional power boosting for \gls{ue} specific \gls{dmrs}), but can also affect \glspl{dci}, e.g., to allow for a ``per transmission'' parameter selection. In addition, \gls{amc} for non-linear systems is a critical missing component that, as we also discuss later in more detail, can determine systems performance. In this context, its absence may be one of the main reasons preventing from adopting non-linear approaches to actual systems.

\section{Evaluation results}

This section presents results obtained by the \gls{ota} measurements that validate our design and provide some indicative performance evaluation of advanced \gls{nl} processing against linear (i.e., \gls{zf}) approaches that serves as the baseline approach for linear processing. Without a loss of generality we employed the soft, near-optimal, non-linear detection algorithm discussed in \cite{nikitopoulos2018massively} and vector-perturbation-based, non-linear precoder introduced in \cite{husmann2018viper} since they are the most promising in terms of processing latency and complexity. 
The number of processing elements assumed are 40 and 32 for uplink and downlink, respectively, that have been observed to provide a good trade-off between error performance and computational complexity. The measurements were conducted using the developed hardware and software \gls{sword} platform for a MU-MIMO setup with 4 and 8-antenna \gls{bs} setup and 4 single antenna \glspl{ue}. While the examined MIMO dimensions are small, and as we will discuss later in detail, they have been sufficient to verify the gains of non-linear processing. It is also significant to note, that the aim of this short evaluation is not only to validate the gains of non-linear approaches compared to linear, but primarily to reveal hidden challenges in the system design that can affect performance.



\noindent\textbf{Measurement setup.}  Several indicative locations were selected for measurements, including three outdoor locations (for uplink measurements only) and four indoor locations (for uplink and downlink measurements). Note that our platform does not currently integrate external power amplifiers. As a result, due to the limited output power, all selected locations had a \gls{los}, and 
the \glspl{snr} was in range of 15dB or below. 
The platform was set in TDD mode, with 5 MHz channel bandwidth and operating frequency of 3.55 GHz. The LTE downlink/uplink slot configuration number 3 which includes 6 downlink slots, 3 uplink slots and 1 special slot was selected for measurements \cite{3gpp.36.211}. A subset of antenna array elements with the same polarisation and equivalent to a \gls{ula} was used. Furthermore, the scheduler for MU-MIMO was set to always schedule transmission to all \glspl{ue} with the same number of resource blocks. \Gls{csi} at the transmitter were obtained using \gls{srs} transmitted in every frame, with a moving average filter applied to reduce the effect of thermal noise. To compensate for any thermal phase drift, each measurement instance was preceded by the TDD reciprocity calibration procedure. 


 \begin{figure}
   \hspace*{1.0cm}\includegraphics[width=0.8\columnwidth]{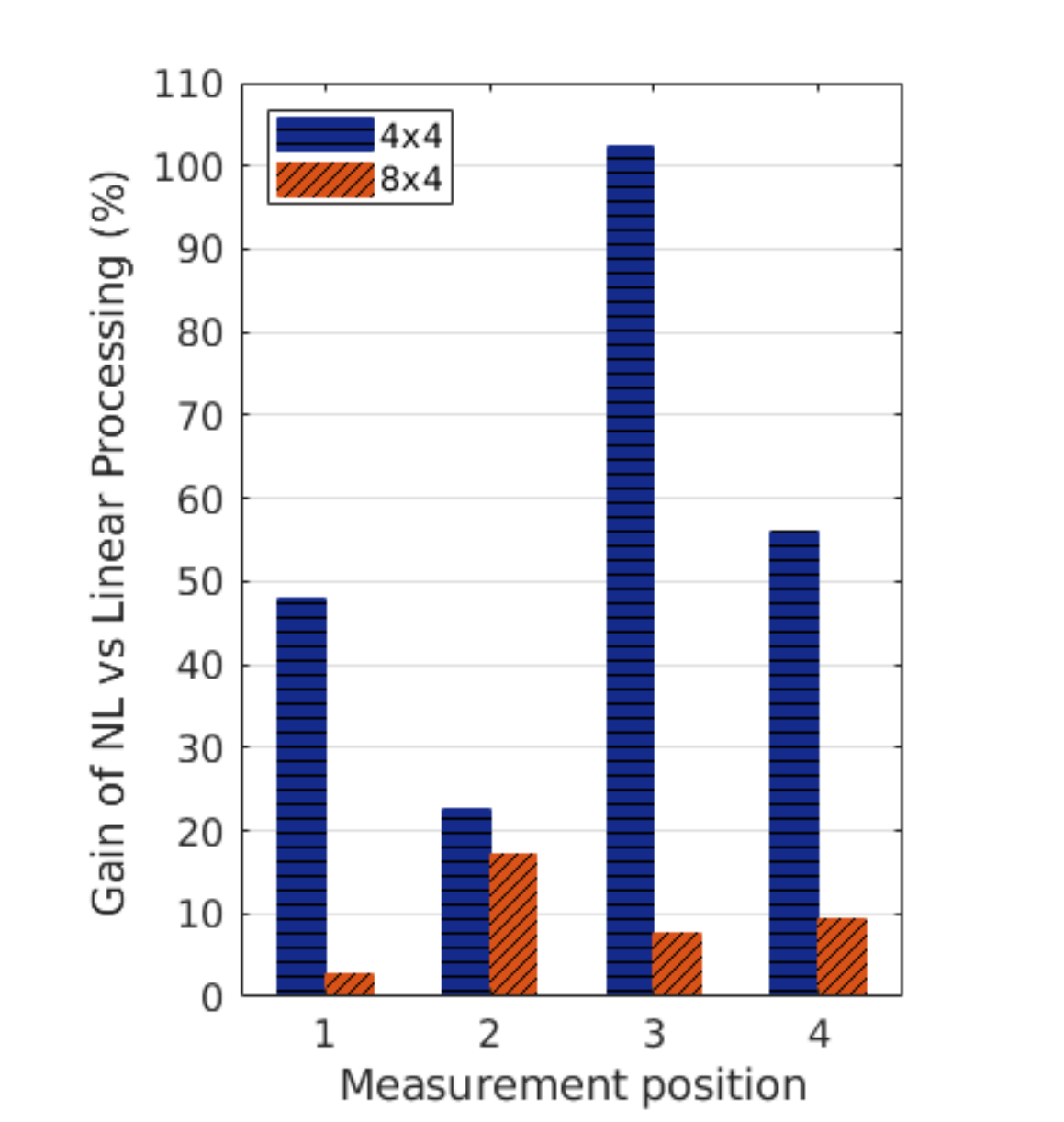}
   \caption{
   Relative gain of NL over Linear processing in uplink for 4x4 and 8x4 MU-MIMO configuration for indoor measurement positions.
   }
   
   \label{fig.indoor.4_4}
   \vspace{-0.4cm}
 \end{figure}

 \begin{figure}
   \hspace*{1.0cm}\includegraphics[width=0.8\columnwidth]{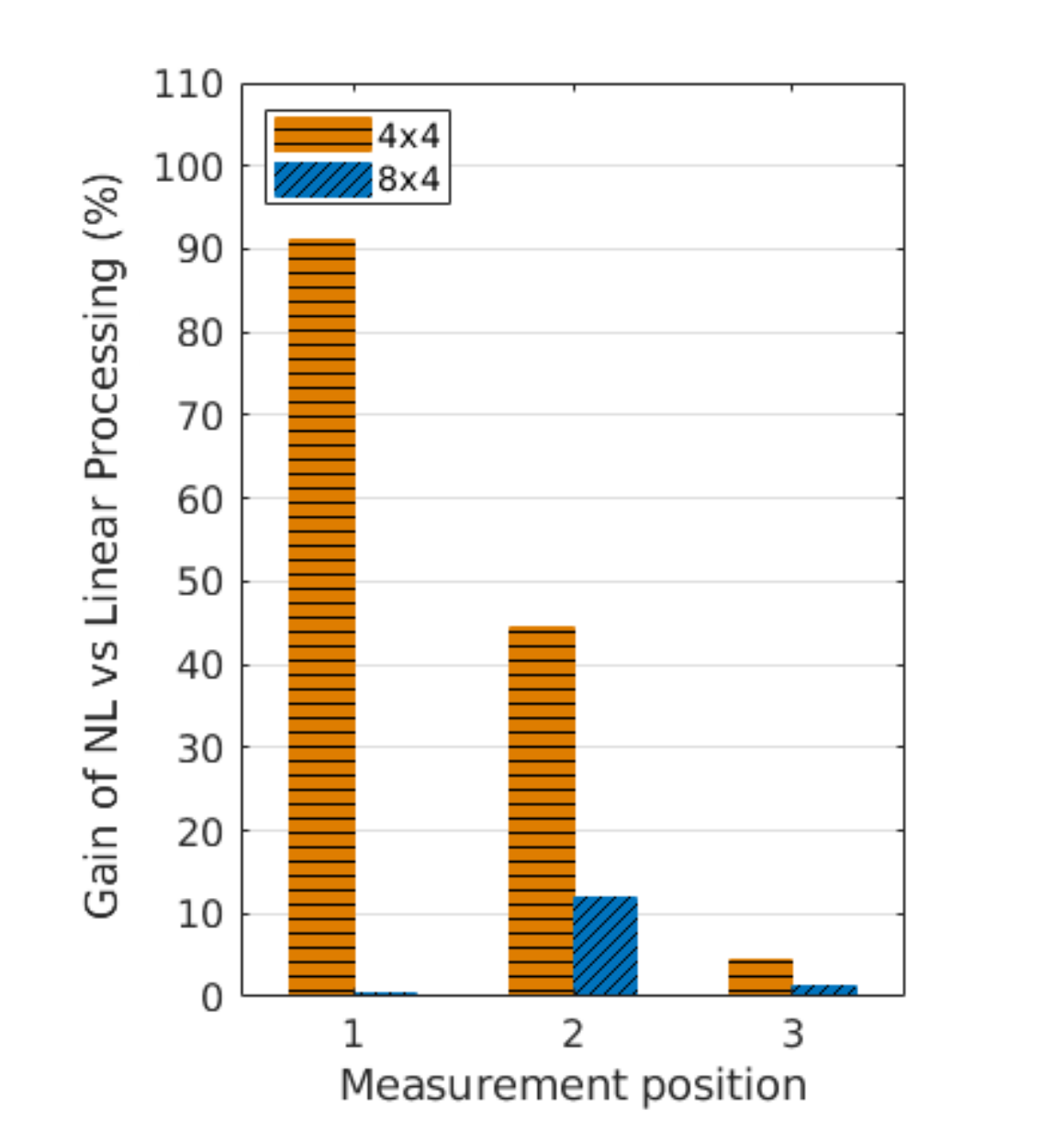}
   \caption{
   Relative gain of NL over Linear processing in uplink for 4x4 and 8x4 MU-MIMO configuration for outdoor measurement positions.
   }
   \label{fig.outdoor.8_4}
   \vspace{-0.4cm}
 \end{figure}

\noindent\textbf{Uplink results.} As can be seen in Figure \ref{fig.indoor.4_4} for our indicative indoor evaluations, and in Figure \ref{fig.outdoor.8_4} for outdoor ones, the use of \gls{nl} detection results in a consistent increase in overall system performance compared to linear. 
In case of the indoor scenario the average gains of NL approaches 57\% and 9\% for 4-antenna, and 8-antenna setup, respectively. In case of the outdoor scenario the average gains of approx. 47\% and 5\% was achieved. The reduced gain in the 8-antenna setup is expected, since increasing the number of BS antennas while maintaining the same number of UEs, allows simplifying the signal detection processing, at the cost of highly under-utilizing the MIMO channel \cite{nikitopoulos2018massively}. Still, in contrast to what has been expected, the gains of the NL processing in the 8x4 MU-MIMO case are non-negligible.


 \begin{figure}
   \hspace*{1.0cm}\includegraphics[width=0.8\columnwidth]{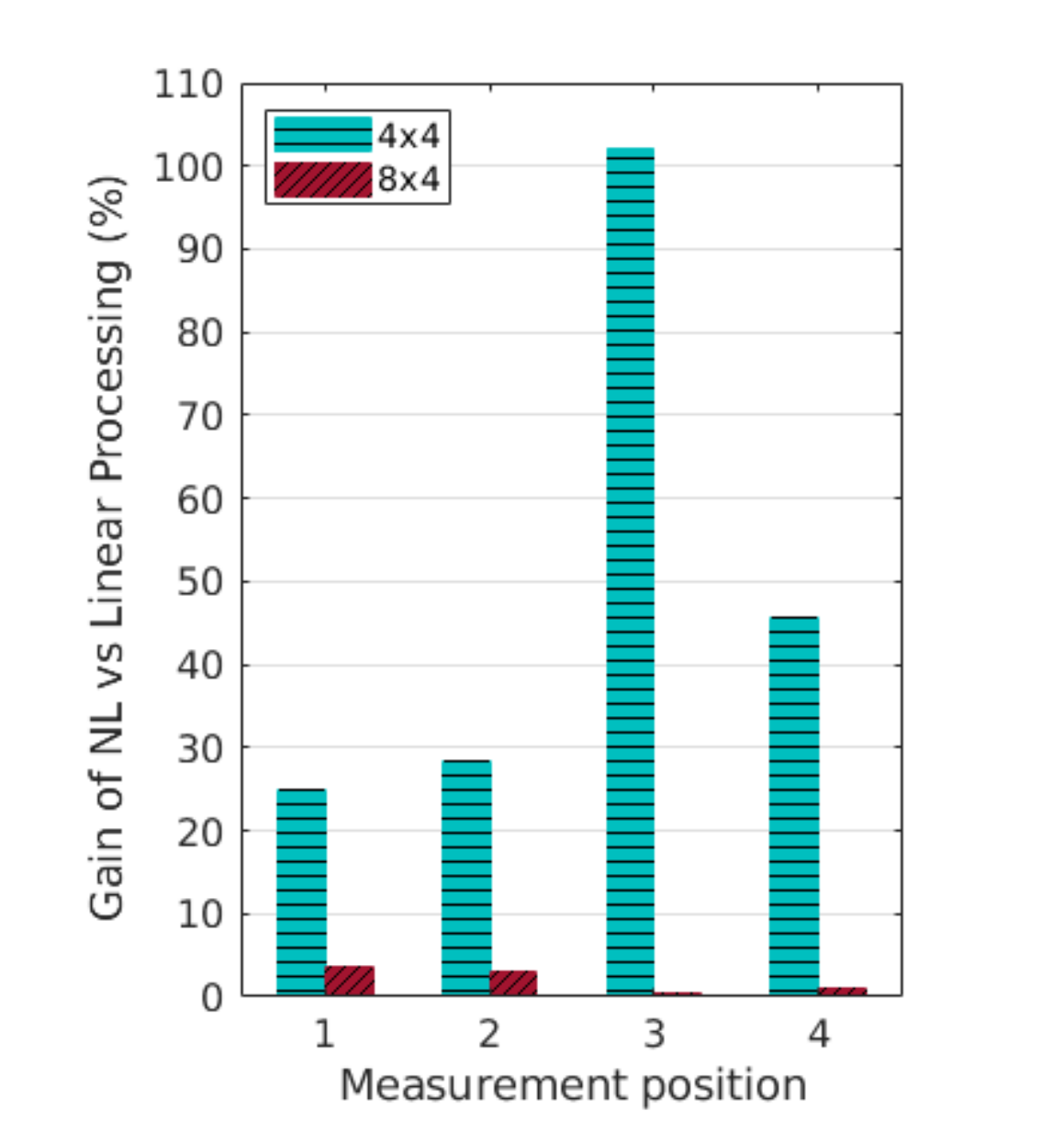}
   \caption{
   Relative gain of NL over Linear processing in downlink for 4x4 and 8x4 MU-MIMO configuration for indoor measurement positions.
   }
   \label{fig.indoor.dl.4_4}
   \vspace{-0.4cm}
 \end{figure}

\noindent\textbf{Downlink results.}
As seen in Figure \ref{fig.indoor.dl.4_4}, similarly to the uplink case, an increase in system performance through the use of NL precoding was obtained in all measured positions, with an average relative gain of approx. 50\%  and 2\% in case of 4x4 and 8x4 MIMO configurations, respectively. While the downlink NL gains compared to linear processing are consistent, they are less prominent compared to the uplink. This is due to reasons like channel aging, as well as because of the imperfections of the SRS-based channel estimation and TDD calibration, that can be further improved.

\section{Remaining Challenges and Way Forward}






Here we discuss some of the remaining challenges that need to be addressed in order to develop future \glspl{bs} that can benefit from the non-linear processing approaches.

\noindent\textbf{Transmission Rate Adaptation.} As it has become obvious, one of the important missing components in order to make non-linear processing both practical and efficient, is how to perform efficient rate adaptation. 
In this direction, two approaches can be potentially examined. The first one is to try to develop \textit{non-linear-specific} \gls{amc} methods, and the other is to adopt \textit{rateless} (or \textit{fountain}) channel encoding.

The direction towards developing non-linear-specific \gls{amc} methods is particularly challenging in the uplink. In this case, the per-user SNR would typically differ and, therefore, each user should transmit using its own transmission rate. This issue could be partially handled by retaining the transmit power control mechanism of \gls{srs} signals (see discussion in Section \ref{remaining_challenges}). Still, the maximum achievable transmission rate is a function of the MIMO channel and the adopted detection method that makes the \gls{amc} problem even more complicated. A promising direction towards non-linear-specific \gls{amc} would be to consider the mathematical framework used for identifying the ``most promising" vector solutions in the massively parallel methods of \cite{nikitopoulos2018massively,FlexCore} since the corresponding \textit{metrics-of-promise} is related to the achievable error-rate probability.
In the downlink, predicting the modulation order and the coding rate that maximizes the throughput could perhaps be easier, since typical non-linear precoding approaches result into the same SNR per user. Still, if a per-user ``power-loading" approach is adopted (that by itself is an interesting research direction) the problem becomes similar to the uplink case, and then the duality between uplink and downlink transmissions could perhaps be explored.  

Instead of a non-linear-specific \gls{amc}, rateless codes can be used, that negate the need for choosing an \gls{mcs} mode \cite{etesamiRaptor,rfc5053,spinal}.  
This is achieved by initially transmitting high-rate information, and then by transmitting parity information, that decreases the effective information rate, until the transmitted information is correctly decoded. Still, such approaches will require revisiting of the way we transmit ACK/NACK signalling. It is noted, that a kin to the rate adoption problem, that is still open and perhaps needs to be considered jointly with \gls{amc}, is the one of ``user-selection'' that allocates users to MIMO transmissions (or MIMO antennas) in order to maximize systems performance.  



\noindent\textbf{Scaling to large numbers of users.}
Channel estimation is an important aspect of every \gls{mimo} system. In order to allow for effective channel estimation in \gls{3gpp} systems, each stream is assigned with a \gls{dmrs} which is orthogonal with respect to \gls{dmrs} allocated for other streams. As \gls{3gpp} systems have not been specifically designed for non-linear processing (which enables supporting very large numbers of users), the number of orthogonal \gls{dmrs} allocations in \gls{3gpp} is limited to 8 in LTE \cite{3gpp.36.211} and 12 in 5G-NR \cite{3gpp.38.300}. Whilst these limits seem reasonable when systems are based on linear processing, for which a number of antennas grows rapidly with a number of streams (making deployment of sites supporting large number of streams impractical), such limits may become a bottleneck in case of systems with non-linear processing (in particular, when number of concurrently supported \glspl{ue} is larger than the number \gls{bs} antennas, as discussed in the next Section).
Note also that \gls{srs} capability to support multiple users, is highly dependent on the channel delay spread and there is only a limited number of cyclic shifts that can be used in practice.
As a result, and given that the \gls{srs} periodicity needs to reflect changes in the channel coherence time, the \gls{srs} capacity may not be sufficient to maintain the \gls{csi} reliability. 

\section{Conclusions \& Future Research}

We have, for the first time, verified in a \gls{3gpp} compliant framework that non-linear processing is a promising approach for increasing the achievable throughput and user connectivity of mobile systems. Still, there is a lot of research yet to be done before such approaches are adopted by actual wireless systems and standards. Among them, two of the most significant open questions are how to perform rate adaption and how to redesign the corresponding wireless systems in order to be able to support a much larger number of users. Especially since, as has already been shown in the literature, the gains of non-linear processing increase when increasing the number of concurrently supported users.

Despite the already verified gains, the most interesting 
capabilities that non-linear \gls{bs} processing can offer, and 
perhaps revolutionize future wireless systems, have not yet been explored. In this direction, we can identify two promising research pathways: (a) non-linear processing for supporting numbers of transmitted information streams that are larger than the number of \gls{bs} antennas, and (b) practical, non-linear, iterative, \gls{bs} processing for further bridging the gap between the theoretical capacity and the achievable throughput of systems with large connectivity. 

\noindent\textbf{Transmitting more streams than base-station antennas.} In a ``fully connected'' wireless ecosystem, future communication systems will need to be able to support a very large number of users. Traditional wireless system designs with linear processing are not capable of supporting more information streams than the number of \gls{bs} antennas and, in practice, can efficiently support only a much smaller number of information streams. Non-linear processing approaches can negate this limitation, at least theoretically, and can promise a supported number of information streams that is much larger than the number of users \cite{GSD,gMultisphere}, even without the need for specifically designated \textit{Non-Orthogonal Multiple Access} (NOMA) techniques \cite{PwrNOMA,LDSOFDM,nikopour2013sparse}. Still, as already discussed, for developing such systems we will need to revisit the signalling procedures, as well as the way we perform channel estimation.

\noindent\textbf{Non-linear, iterative, base-station processing.} Iterative systems that exchange ``soft'' information between a non-linear detector and a ``soft'' channel decoder promise substantial gains \cite{hochwald2003achieving,STSsoftin,LLRclip}. Still, such approaches are not scalable to large number of information streams due to their exponentially increased complexity and latency requirements. For example, the approximate non-linear approach of \cite{LLRclip} would require a  $10^{14}$ multiplications for a $12\times12$ MIMO system. On the other hand, currently proposed massively parallel, soft-output approaches that can substantially reduce the corresponding complexity and processing latency requirements \cite{nikitopoulos2018massively,gMultisphere} are not applicable to the iterative case. This is because such approaches are heavily based on the geometrical properties of the transmitted signal constellation which is destroyed by the existence of prior (from previous iterations) information.  Furthermore, existing iterative schemes, cannot currently support a larger number of information streams than the \gls{bs} antennas. Developing non-linear, massively-parallel, iterative detection, decoding techniques, able to support more users than \gls{bs} antennas, could give a significant connectivity boost, and allows us to access unexploited capacity resources. 


\bibliographystyle{IEEEtran}
\bibliography{main}

 \small
 \printglossary[type=\acronymtype,title=List of Acronyms]

\end{document}